\documentclass[%
 amsmath,amssymb,
 graphicx,
 reprint]{revtex4-1}

\draft 

\usepackage{graphicx}
\usepackage{dcolumn}
\usepackage{bm}

\usepackage[utf8]{inputenc}
\usepackage[T1]{fontenc}
\usepackage{mathptmx}
\usepackage{etoolbox}
\usepackage{mathtools}
\usepackage{textcomp}

\makeatletter
\def\@email#1#2{%
 \endgroup
 \patchcmd{\titleblock@produce}
  {\frontmatter@RRAPformat}
  {\frontmatter@RRAPformat{\produce@RRAP{*#1\href{mailto:#2}{#2}}}\frontmatter@RRAPformat}
  {}{}
}%

\begin{document}

\title{Multi-line lasing in the broadly tunable ammonia quantum cascade laser pumped molecular laser}



\author{Paul Chevalier}
\author{Arman Amirzhan}
\affiliation{Harvard John A. Paulson School of Engineering and Applied Sciences, Harvard University, Cambridge, MA 02138, USA}
\author{Jeremy Rowlette}
\author{H. Ted Stinson}
\author{Michael Pushkarsky}
\author{Timothy Day}
\affiliation{DRS Daylight Solutions, San Diego, CA 92128, USA}
\author{Federico Capasso}
\affiliation{Harvard John A. Paulson School of Engineering and Applied Sciences, Harvard University, Cambridge, MA 02138, USA}
\author{Henry O. Everitt}
\email[Authors to whom correspondence should be addressed: ]{capasso@seas.harvard.edu (F.C) or everitt@phy.duke.edu (H.O.E.)}
\affiliation{DEVCOM Army Research Lab, Houston, TX  77005, USA}
\affiliation{Department of Physics, Duke University, Durham, NC 27708, USA}





\begin{abstract} 
Gaseous ammonia has previously been demonstrated as a compelling gain medium for a quantum cascade laser pumped molecular laser (QPML), exhibiting good power efficiency. Here we explore the potential of the ammonia QPML to produce powerful, broadly tunable terahertz frequency lasing on rotational and pure inversion transitions. After theoretically predicting possible laser frequencies, pump thresholds, and efficiencies, we experimentally demonstrate unprecedented tunability -- from 0.763 to 4.459 THz  -- by pumping Q- and R-branch infrared transitions with widely tunable quantum cascade lasers. We additionally demonstrate two types of multi-line lasing: simultaneous pure inversion and rotation-inversion transitions from the same pumped rotational state, and cascaded lasing involving transitions below the pumped rotational state. We report single frequency power levels as great as 0.45 mW from a low volume laser cavity.
\end{abstract}

\maketitle

The quest for powerful, tunable, continuous wave sources of terahertz radiation for use in
spectroscopy, communications, imaging, and radar has recently been accelerated by the
development of a new type of source: a quantum cascade laser-pumped molecular laser
(QPML).\cite{chevalier2019widely,pagies2016low,lampin2020quantum}
A molecular gas in a compact laser cavity can be made to lase on any rotational
transition of virtually any molecule that has a permanent electric dipole moment, a vapor
pressure, and an infrared (IR) rotational-vibrational absorption band that may be spanned by a tunable quantum cascade laser (QCL).\cite{chevalier2019widely} By tuning the QCL frequency to coincide with any such ro-vibrational transition $J_L \to J_U$, it has been previously shown that two rotational population inversions may be created, the direct transition between $J_U \to J_U -1$ of the excited vibrational level and the refilling transition between $J_L + 1 \to  J_L$ of the ground vibrational level.~\cite{chevalier2019widely,wang2018high} The length of the
cavity containing the gas may then be adjusted so one of its resonant frequencies coincides with one of these transitions. Thus, a single molecular gain medium may be made to lase simply by tuning both the QCL to a specific IR frequency and the QPML cavity to the terahertz frequency of the transition. In this manner, any molecular rotational absorption transition may be made to lase, and millions of such transition frequencies have been measured or tabulated.\cite{JPL2017,Splatalogue2007,HITRAN2020}

Ammonia was the first molecule made to lase in this manner,~\cite{pagies2016low} although not on rotational transitions but on widely-spaced pure inversion transitions in the $\nu_2=1$ excited vibrational band. 
The pure inversion transition frequencies in $\nu_2 = 1$ are much larger than the well-known ground state microwave transitions (9-35 GHz) because the inversion energy is comparable to the height of the tunnel barrier. Consequently, the inversion splittings in $\nu_2 = 1$ are strong transitions clustered near 1.1 THz. A handful of such pure inversion laser transitions have been reported, including one that yielded 1~mW of output power, when a Q-branch ($J_L = J_U = J$) transition was pumped by a narrowly tunable QCL operating between 965-968 cm$^{-1}$.~\cite{lampin2020quantum} 
More recently, additional lasing transitions pumped by IR transitions in NH$_3$ have been reported.~\cite{wienold2020laser,mammez2022optically}


By using a widely tunable pump source, we have demonstrated that the simple linear molecule nitrous oxide (N$_2$O) can achieve continuous-wave laser emission for at least 39 transitions with discrete line tunability from 0.25 to 0.95 THz.~\cite{chevalier2019widely}  More recently we demonstrated that using the prolate symmetric top molecule methyl fluoride (CH$_3$F) as the gain medium allows for even greater tunability with emission on at least 120 lines spanning 0.25 to 1.3 THz.~\cite{amirzhan2022quantum}

In this letter, we demonstrate the broad tunability that is possible from the ammonia QPML when rotation-inversion transitions are made to lase. We have discovered and characterized two multi-line lasing mechanisms in the ammonia QPML: the simultaneous lasing of a pure inversion and a rotation-inversion transition sharing the same pumped upper level, and the cascaded lasing from a lower energy lasing transition indirectly connected to the pumped level. Both mechanisms allow simultaneous lasing at two very different frequencies, and in certain cases the cascade mechanism allows for lasing at frequencies that would not otherwise be possible due to the selection rules.

\begin{figure}[htbp]
\centering
\includegraphics[width=\linewidth]{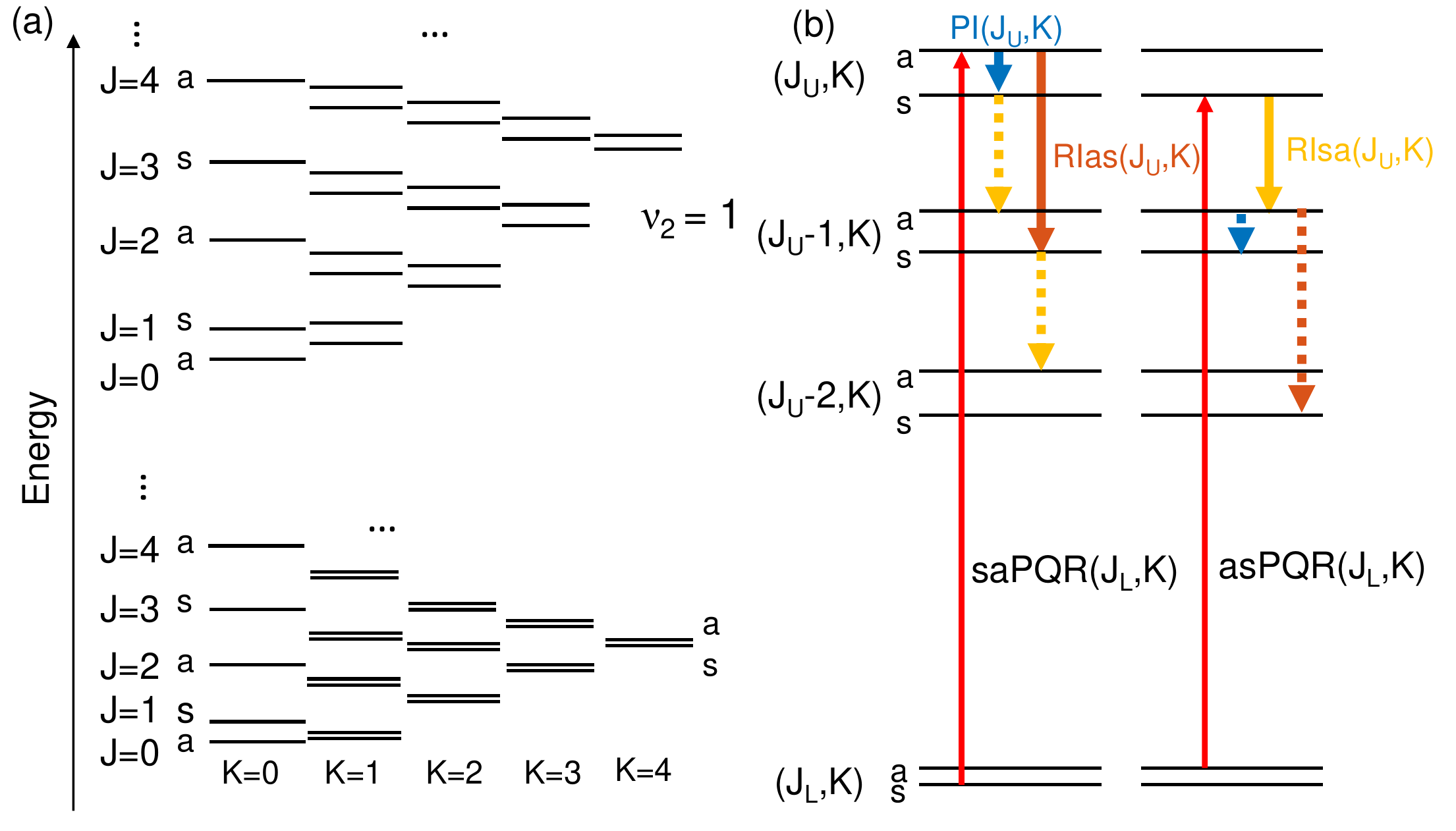}
\caption{(a) Energy level diagram of NH$_3$ in the ground and $\nu_2=1$ vibrational states. Note the alternating s and a levels at $K=0$. (b) Three types of laser transitions are possible in NH$_3$ QPMLs: pure inversion (PI, blue) and rotation-inversion (RIas, red) when an s $\to$ a IR transition is pumped, and rotation-inversion (RIsa, yellow) for an a $\to$ s IR pump.  Cascaded transitions for each type are illustrated as dashed arrows of the same colors.
}
\label{fig:fig1}
\end{figure}

It is well known that ammonia’s rotational spectrum is complicated by the inversion splitting caused by the inversion tunneling of the nitrogen atom in this pyramidal molecule\cite{gordy1984microwave,townes1975microwave}. A simplified version of the energy level diagram for the ammonia molecule for $J\le 3$ is given in Fig.~\ref{fig:fig1}(a), where the $J$ is the quantum number representing the total angular momentum, and $K$ represents the projection of the angular moment along the symmetry axis of the molecule. Whether for pumping or lasing, the selection rules require a change in symmetry of the two energy levels associated with the inversion doublet, symmetric (s) and anti-symmetric (a), so the only allowed transitions are a $\to$ s or s $\to$ a. Consequently, pure inversion transitions and Q-branch transitions ($\Delta J=0$) with $K = 0$ are forbidden. Because NH$_3$ is an oblate symmetric top, the lowest energy levels occur for $J = K$, and these are generally the most heavily populated.

Because of the inversion splitting, two types of P, Q, or R branch IR transitions may pump population into a given rotational state $(J,K)$ within $\nu_2=1$ (see Fig.~\ref{fig:fig1}(b)). The first (left side) is a higher energy transition connecting a lower symmetric inversion level in the ground vibrational state to a higher antisymmetric inversion level in $\nu_2 = 1$, denoted saP$(J,K)$, saQ$(J,K)$, or saR$(J,K)$, respectively spanning 679-948, 954-968, or 1007-1249 cm$^{-1}$ for $0 \le J\le 14$. The converse transitions asP$(J,K)$, asQ$(J,K)$, and asR$(J,K)$ (right side) occur at lower frequencies for a given J (respectively 632-932, 905-949, or 952-1237 cm$^{-1}$ for $0 \le J\le 14$).

The selection rules for a lasing transition are $\Delta J = 0$, $\pm1$, $\Delta K = 0$, $s \leftrightarrow a$. Although ``forbidden’’ transitions with $\Delta K$=3, 6, 9... may be  allowed,~\cite{belov1980inversion,oka2021forbidden} their threshold for lasing will be higher than the $\Delta K$=0 lines considered here. When an saQ$(J,K)$ transition is pumped, two QPML direct lasing transitions may occur (shown on the left side of Fig.~\ref{fig:fig1}(b)): the pure inversion transition $(J,K,a)$ $\to$ $(J,K,s)$ and the rotation-inversion transition $(J,K,a)$ $\to$ $(J-1,K,s)$. Because of symmetry-based selection rules, only sa pump transitions can directly produce lasing on pure inversion transitions. When an asQ$(J,K)$ transition is pumped, the rotation-inversion transition $(J,K,s)$ $\to$ $(J-1,K,a)$ may be made to lase (right side of Fig.~\ref{fig:fig1}(b)). For simplicity, we label these three QPML transitions in $\nu_2=1$ as PI$(J,K)$, RIas$(J,K)$, and RIsa$(J,K)$, respectively. Similarly, pumping an R$(J,K)$ or P$(J,K)$ transition may induce lasing on transitions with $J_U = J+1$ or $J_U = J-1$, respectively. In the following, we consider all three types of lasing transitions from Q or R branch pump transitions, both s $\to$ a and a $\to$ s. 


\begin{figure}[htbp]
\centering
\includegraphics[width=\linewidth]{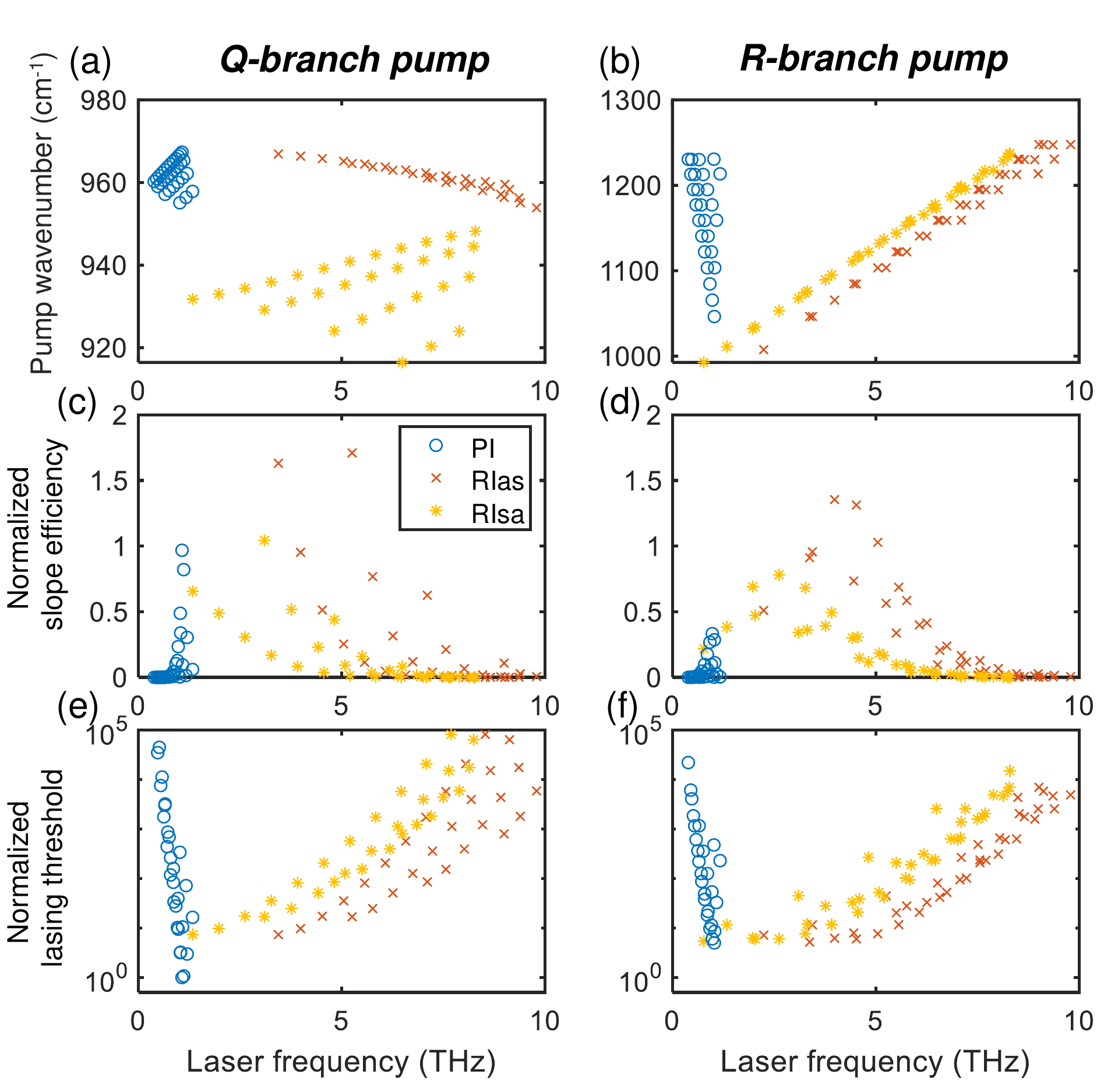}
\caption{Direct lasing transition frequencies for the NH$_3$ QPML, plotting only those transitions for which $K$ is a multiple of 3, for (a) a Q-branch or (b) an R-branch pump. Normalized slope efficiency (c,d) and threshold (e,f) of the NH$_3$ QPML for those transitions, using the method of Ref.~\citenum{amirzhan2022quantum} for a Q-branch (c,e) or an R-branch (d,f) pump.}
\label{fig:fig2}
\end{figure}

To explore the relative tunability and efficacy of all three types of transitions systematically, we first compute the energy levels of ammonia as a function of $J$, $K$, and inversion state (a or s) using the published rotational constants for ammonia.
\cite{yu2010submillimeter} From these energy levels and the selection rules mentioned above, one can compute the expected direct lasing frequencies as a function of the pump wavenumber. For Q and R-branch IR pump transitions, QPML frequencies for representative transitions (those for which $K$ is a multiple of 3) are plotted respectively in Figs.~\ref{fig:fig2}(a) and ~\ref{fig:fig2}(b). These plots reveal that ammonia can be used as a gain medium in a widely tunable QPML spanning the frequency range 0.140 – 9.634 THz, far beyond the limited tunability afforded by PI transitions clustered at 1.1 $\pm$ 0.3 THz.

But how much power can be expected from them?  To understand how power depends on $(J,K)$, we use the method described in our previous work~\cite{amirzhan2022quantum} to compute the normalized slope efficiency and lasing threshold power for direct PI and RI lasing transitions of the ammonia laser. By eliminating the factors that depend on cavity geometry, pump power, and gas pressure, this calculation provides unitless quantities that allow line-by-line comparisons of threshold and slope efficiency.  The method requires computation of the population fraction for each energy level and the branching ratios for both the pumped transition (P-, Q-, or R-branch) and the corresponding lasing transitions (PI, RIas, and RIsa) as a function of $J$, $K$, and inversion state.~\cite{gordy1984microwave} Q-branch and pure inversion transitions are governed by a branching ratio that goes as $K^2$, while the R-branch and rotation-inversion transitions are governed by branching ratios that go as $(J+1)^2 - K^2$ or $J^2 - K^2$, respectively. Thus, Q-branch and pure inversion transitions are favored for large $K$, while R-branch and rotation-inversion transitions are favored for small $K$. 

For Q-branch-pumped lasing transitions whose $K$ is a multiple of 3, Figures~\ref{fig:fig2}(c) and~\ref{fig:fig2}(e) respectively give the slope efficiency and lasing threshold, normalized by the values for the lowest threshold line (pure inversion $J=4,K=3$). Similar plots are shown in Figs.~\ref{fig:fig2}(D) and~\ref{fig:fig2}(f) for an R-branch pump using the same normalization. With the exception of a few PI transitions, the transitions with the lowest threshold and highest slope efficiencies are RI transitions between 1 and 6 THz. The supplementary information (SI) for this manuscript includes a detailed explanation of the normalization process, followed by complete plots for all transitions and a table giving all the transition frequencies for $J_L \le 10$.

\begin{figure}[htbp]
\centering
\includegraphics[width=\linewidth]{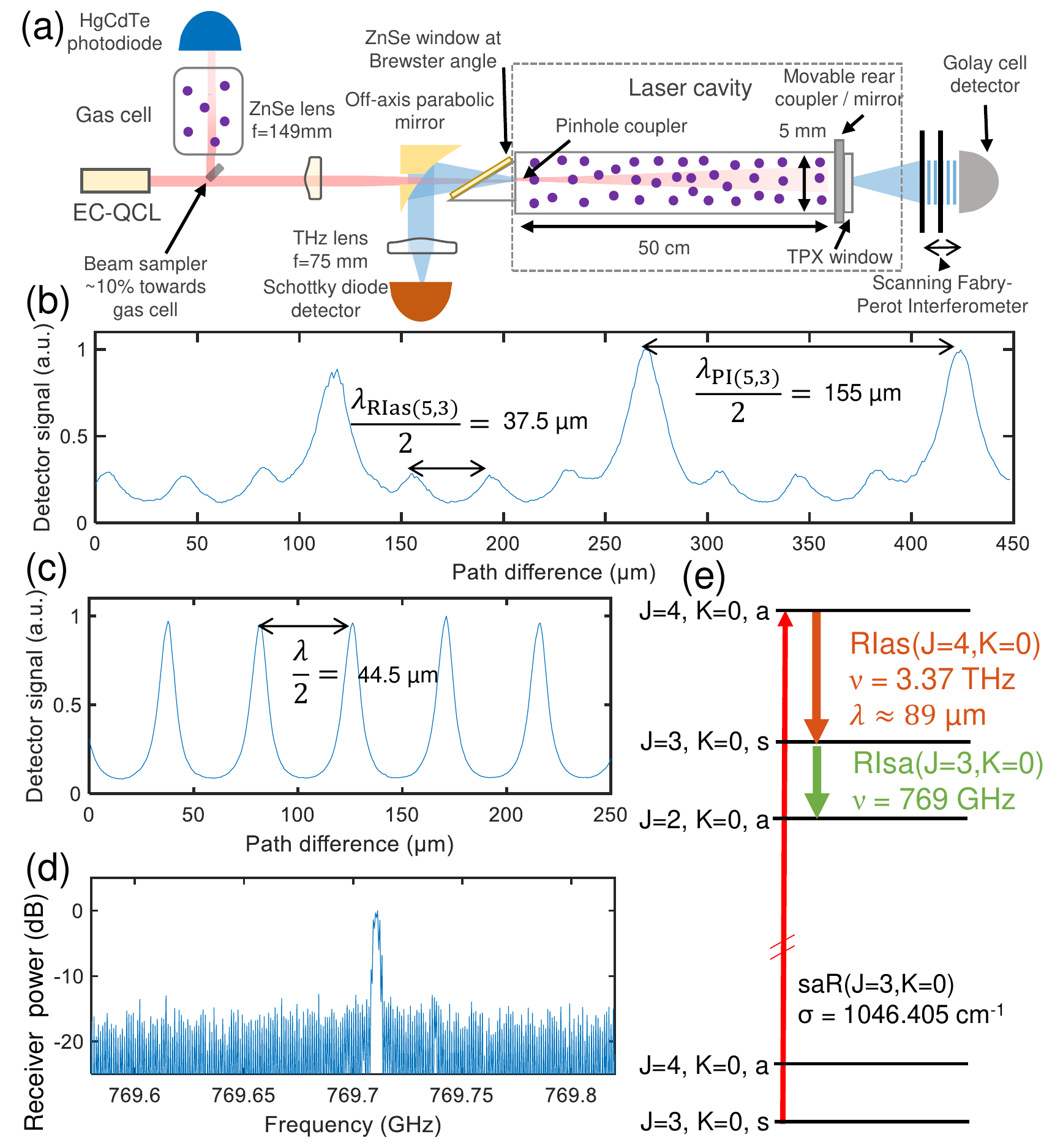}
\caption{(a) Experimental schematic of the NH$_3$ QPML: the 4.8 mm diameter, 50 cm long copper laser cavity has pinhole output couplers in the front and rear.  (b) Interferogram measured with a scanning Fabry Perot inteferometer (SFPI), showing simultaneous lasing at $\sim$0.98 THz and $\sim$3.99 THz (wavelengths 310 \textmu m and 75 \textmu m). (c) Measured interferogram showing a laser emission frequency of  $\sim$3.37 THz (89 \textmu m wavelength). (d) Signal measured with a receiving mixer showing laser emission at 769.710 GHz. (e) Mechanism for simultaneous direct and cascaded laser emission in a NH$_3$ QPML pumped by the asR$(3,0)$ IR transition. }
\label{fig:fig3}
\end{figure}

To verify these predictions, we used the experimental setup depicted in Fig.~\ref{fig:fig3}(a). IR light from an external cavity (EC)-QCL (Daylight Solutions 41095-HHG-UT, tunable from 920 to 1194 cm$^{-1}$) is injected through a Brewster's angle ZnSe window into a THz cavity that consists of a 50 cm long copper pipe with a 4.8 mm internal diameter. A flat mirror in the front of the cavity with a centered 1 mm diameter pinhole is used as an output coupler. The cavity resonance frequency was tuned by changing the rear tuning mirror position to adjust the cavity length. A separate gas absorption cell containing 100 mTorr of anhydrous ammonia was used to monitor how precisely the QCL emission frequency was tuned to the desired NH$_3$ ro-vibrational transition.

Unlike previous demonstrations of the N$_2$O and CH$_3$F QPMLs, the rear mirror of the cavity can either be a flat mirror or a pinhole coupler (pinhole diameter 1.7 mm) sealed by a 4 mm thick polymethylpentene (TPX) window. The front ZnSe coupler absorbs light at frequencies above 2 THz, thereby acting as a low pass terahertz filter, while the TPX rear window transmits all terahertz frequencies. Outgoing terahertz light generated in the laser cavity can be collected from the front pinhole of the laser by an off-axis parabolic mirror and focused onto a detector using a 75 mm focal length TPX lens.  However, output power was greater when measured from the rear coupler with the larger pinhole and was not affected by the frequency dependent response of ZnSe Brewster window or the hole in the off-axis parabolic mirror. Light from the rear of the cavity was characterized by a Tydex TSFPI-2 scanning Fabry-Perot interferometer (SFPI) coupled with a broadband THz detector (Golay cell). The periodicity of interference fringes as a function of the path difference corresponds to the half-wavelength of the emitted light.

We pumped the QPML cell containing 20 mTorr of ammonia with the widely tunable QCL and measured 24 direct PI and RI lasing transitions spanning 0.769710-4.458880 THz by tuning the QCL across Q- and R-branch transitions from 966.269-967.407 and 992.699-1084.620 cm$^{-1}$. As can be seen in Fig.~\ref{fig:fig1}(b), either a pure inversion (blue solid arrow) or a rotation-inversion (red solid arrow) transition may be made to lase.  Surprisingly, when we pumped the saR$(4,3)$ IR transition at 1065.565 cm$^{-1}$, we observed both PI and RI lines to lase simultaneously. Figure~\ref{fig:fig3}(b), which plots the measured signal as a function of the path difference of the SFPI, exhibits two periodicities corresponding to simultaneous emission near 0.980 THz (PI$(5,3)$) and 3.99 THz (RIas$(5,3)$). Without a spectrometer to discriminate these two lines, the total power emitted could have been mistakenly ascribed to just one of them. Experimentally, the relative magnitude of each line can be adjusted by changing the cavity length and thereby tuning the cavity resonance to match either frequency.

We then pumped the same cell with the QCL tuned to the saR$(3,0)$ IR transition (1046.405 cm$^{-1}$), since the Q-branch version of this transition is forbidden. The pure inversion lasing transition is also forbidden for $K=0$ because the (4,0,s) and (3,0,a) levels do not exist. The QPML could only lase on the RI transition, and light from that RIas(4,0) transition at 3.3736 THz was measured from the rear output coupler (solid red arrow in Fig.~\ref{fig:fig1}(b)). The emission frequency was measured by the SFPI, using the interferogram shown in Fig.~\ref{fig:fig3}(c).

Surprisingly, light was also measured from the front output coupler at 769.710 GHz, using a receiving mixer (VDI SAX WR1.0), as shown in Fig.~\ref{fig:fig3}(d). According to the selection rules, calculated energy levels, and published transition frequencies for ammonia\cite{Splatalogue2007,JPL2017,HITRAN2020,yu2010submillimeter,pearson2016modeling} only direct lasing at 3.373613 THz should be expected. The presence of light at two significantly different frequencies for the saR$(3,0)$ pump transition can be explained by the energy diagrams in Figs.~\ref{fig:fig1}(b) (yellow dashed arrow below solid red arrow) and \ref{fig:fig3}(e). Molecules that mediated the laser emission at 3.373613 THz (RIas$(4,0)$) have relaxed to (3,0,s). At low gas pressures where collisional relaxation is very slow, these molecules in (3,0,s) created another population inversion above (2,0,a). This second population inversion can then emit light at the RIsa$(3,0)$ frequency once it reaches the lasing threshold, which is what was observed. Note that this cascaded transition, measured at 769.710 GHz, is not the diagonal transition near 772.595 GHz caused by an accidental $\Delta$K = 3 degeneracy first reported by S.P. Belov et al.~\cite{belov1980inversion}.

Unlike the situation shown in Fig.~\ref{fig:fig3}(b), the SFPI interferogram in Fig.~\ref{fig:fig3}(c) does not a exhibit a second periodicity at half the wavelength of the cascaded line (195 \textmu m). This shows that the power emitted at the cascaded line is much lower than at the direct frequency. We measured the output power of the laser emitting around 3.37 THz using a calibrated calorimeter detector (VDI PM5B) combined with a conical horn matched with the waveguide input of the power meter. We placed the conical horn right in front of the rear TPX window and measured an output power as large as 0.45 mW in this configuration. In the SI we plot the measured beam profile. Because of its magnitude, we can attribute at least 95\% of this output power to the direct line lasing at 3.373613 THz. After accounting for the Manley-Rowe factor,\cite{chevalier2019widely} the internal quantum efficiency for converting IR photons into THz photons is approximately 2.8\% for this transition, and may be as large as 11\% (1.8 mW output) if absorption and reflection from the collection optics introduce as much as a factor of four loss from emission to detection.

\begin{figure}[htbp]
\centering
\includegraphics[width=\linewidth]{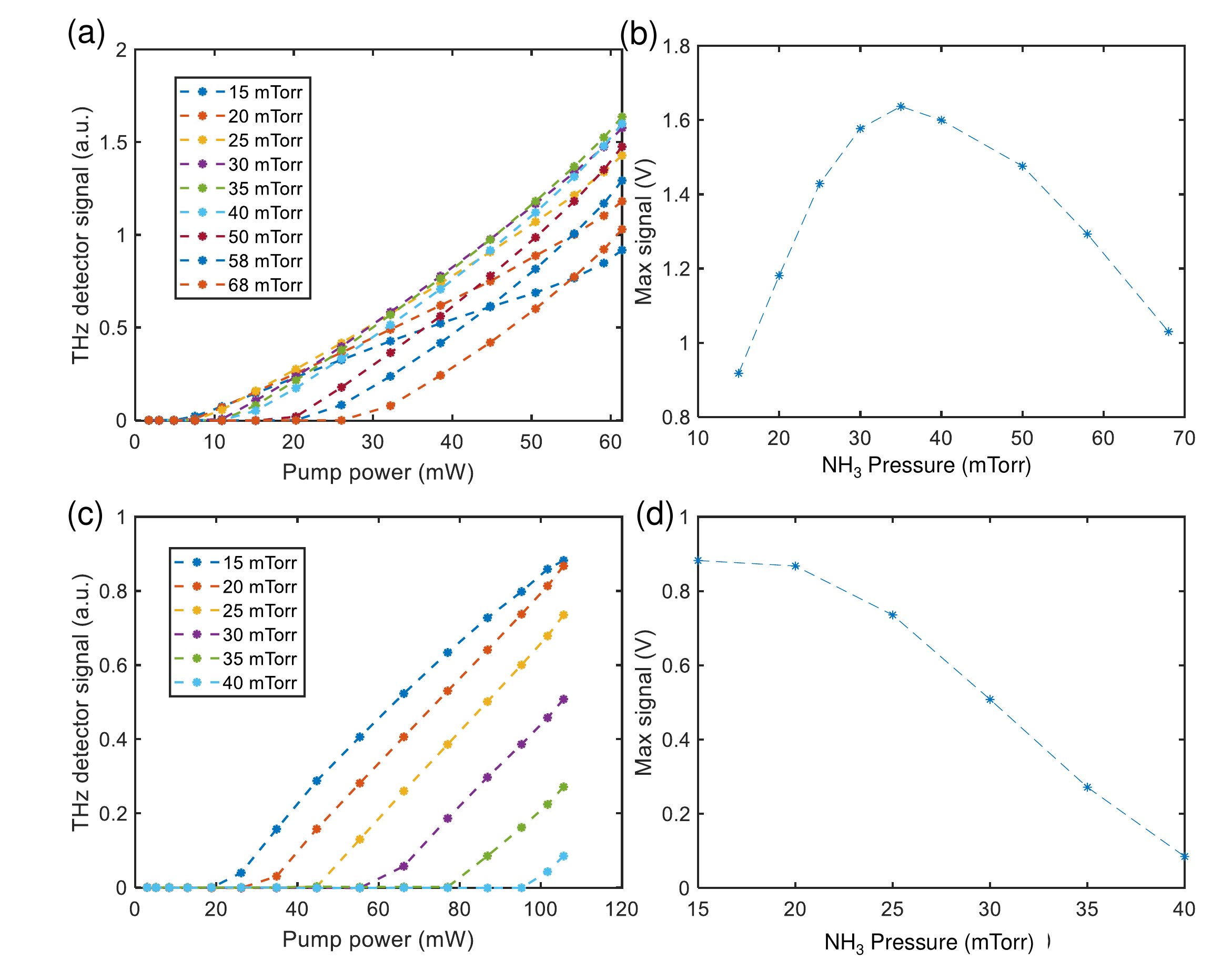}
\caption{(a) Plot of the detected signal at 769.710 GHz for an infrared pump at 992.699  cm$^{-1}$ (direct lasing line) as a function of the pump power and gas pressure. (b) Plot of the peak signal from (a) as a function of the gas pressure. (c) Plot of the detector signal at 769.710 GHz for an infrared pump at 1046.405 cm$^{-1}$ (cascaded lasing line) as a function of the pump power and gas pressure. (d) Plot of the peak signal from (c) as a function of the gas pressure.}
\label{fig:fig4}
\end{figure}

\begin{table*}[htp]
\caption{\label{tab:measuredCascadedLines} Table giving the frequency of measured cascaded lines along with their direct pump wavenumber.}
\begin{tabular*}{0.65\textwidth}{c|c|c|c|c}
\hline\hline
Frequency  & Transition & Direct R pump & Cascaded R pump & Direct Q pump \\
(GHz) & type & wavenumber (cm$^{-1}$) & wavenumber (cm$^{-1}$) & wavenumber (cm$^{-1}$) \\
\hline
769.710	& RIsa(3,0) & asR(2,0) @ 992.699 & saR(3,0) @ 1046.405 & not allowed \\
762.852 & RIsa(3,1) & asR(2,1) @ 992.450 & saR(3,1) @ 1046.401 & asQ(3,1) @ 967.449 \\
	&  &  & or saR(2,1) @ 1027.047 & \\
1393.079 & RIsa(4,1) & asR(3,1) @ 1013.176 & saR(4,1) @ 1065.594 & asQ(4,1) @ 967.031 \\
	&  &  & or saR(3,1) @ 1046.401 & \\
1073.050 &	PI(3,3) & not allowed & asR(3,3) @ 1011.204 & asQ(3,3) @ 967.346\\
1030.530 & 	PI(4,3) & saR(3,3) @ 1046.375	 & asR(4,3) @ 1032.131 & asQ(4,3) @ 966.905 \\
979.650 &	PI(5,3) & saR(4,3) @ 1065.565	& asR(5,3) @ 1053.130 & asQ(5,3) @ 966.380 \\
\hline
\end{tabular*}
\end{table*}

In order to confirm cascaded lasing, we measured the detected signal at 769.710 GHz as a function of pump power and ammonia pressure for both the direct RI transition (pumped by asR$(2,0)$ at 992.699 cm$^{-1}$) and the cascaded line (pumped by saR$(3,0)$), and the result is plotted in Figs.~\ref{fig:fig4}(a) and~\ref{fig:fig4}(c) respectively. These measurements were both conducted using a flat mirror instead of a pinhole coupler at the rear of the cavity in order to lower the radiative cavity losses. Although the output power obtained through the front coupler was significantly reduced, we nevertheless observe that the cascaded lasing threshold is higher than the direct lasing threshold, an indication of different excitation mechanisms. 

As further confirmation of the different excitation mechanisms, we measured the peak signal as a function of NH$_3$ gas pressure using the maximum available pump power for the respective pump frequencies (89 mW at 992.699 cm$^{-1}$, 152 mW at 1046.405 cm$^{-1}$). Figures~\ref{fig:fig4}(b) and~\ref{fig:fig4}(d) reveal that cascaded lasing is more efficient at low pressure, reaching its maximum power at 15 mTorr, while direct lasing is more efficient at higher pressure, reaching its maximum power at 30 mTorr.

This \textit{cascaded} lasing, as opposed to \textit{direct} lasing, has been seen on a few transitions in other optically pumped far-IR laser gases at low pressures and high pump powers.\cite{chang1970cw,button2013reviews} However, with tunable QCLs this cascaded lasing at low pressures is possible on most transitions in most molecules, including ammonia. As summarized in Table I, we observed cascaded lasing on six lines, including lines for which $K\neq 0$ so that PI transitions are also allowed. For example, when pumping on the saR$(3,1)$ IR transition, we observed both cascaded lasing at 762.852 GHz, corresponding to the RIsa$(3,1)$ line, and direct lasing at 973.827 GHz, corresponding to the PI$(4,1)$ line, just by appropriately tuning the cavity resonance. Because of the low $K$ value, the threshold of the PI line was larger than the threshold of the cascaded transition. In general, the choice of the $K$ value is particularly important to determine which type of cascaded line can be made to lase. As noted above, pure inversion transitions are favored for large $K$ while rotation-inversion transitions are favored for small $K$. Thus, by choosing $K$ close to 0, the threshold of any pure inversion line is significantly increased, so direct or cascaded lasing is more likely to happen on a rotation inversion line.
Conversely, by choosing $K$ close to $J$, PI transitions will have a lower lasing threshold and will be more likely to be observed as direct or cascaded transitions.

From these observations it has become apparent that two types of simultaneous multi-line lasing transitions -- multiple direct and direct-cascaded -- exist in ammonia (see Fig.~\ref{fig:fig1}(b)).  Specifically, the saPQR$(J_L,K)$ pump can generate simultaneous lasing on two direct transitions, PI$(J_U,K)$ and RIas$(J_U,K)$, each of which can produce lasing on a different cascaded laser transition, RIsa$(J_U,K)$ and RIsa$(J_U-1,K)$ respectively.  By contrast, the asPQR$(J_L,K)$ pump can create only one direct transition, RIas$(J_U,K)$, but it can produce two cascaded lasing transitions, PI$(J_U-1,K)$ and  RIas$(J_U-1,K)$.  In either case, which lines actually lase is determined by the transition-unique branching ratios, energy levels, and cavity tuning, but it is possible that up to four or three lines of significantly different frequency may lase simultaneously in the former or latter cases, respectively. In all, 34 lasing transitions were observed, spanning 0.762852 to 4.458880 THz (see SI for details), including seven that were observed in both direct and cascaded modes. This multi-line lasing attribute adds further excitement to the ammonia QPML, whose unique energy level spectrum permits broad single line tunability from 0.140 to at least 9.634 THz.

\begin{acknowledgments}
The authors acknowledge S. Cotreau and A. DiMambro of Harvard University Instructional machine shop for their help with fabrication of the THz cavity elements. This work was partially supported by the U.S. Army Research Office (contracts W911NF-19-2-0168, W911NF-20-1-0157) and DRS Daylight Solutions. Any opinions, findings, conclusions, or recommendations expressed in this material are those of the authors and do not necessarily reflect the views of the Assistant Secretary of Defense for Research.
\end{acknowledgments}

\section*{Supplementary Material}
See Supplement 1 for additional details about the experimental and theoretical work presented in this manuscript.

\section*{Data Availability Statement}
The data that support the findings of this study are available from the corresponding author upon reasonable request.

\section*{References}
\bibliography{references}

\end{document}


\title{Supplementary document: Multi-line lasing in the broadly tunable ammonia quantum cascade laser pumped molecular laser}



\author{Paul Chevalier}
\author{Arman Amirzhan}
\affiliation{Harvard John A. Paulson School of Engineering and Applied Sciences, Harvard University, Cambridge, MA 02138, USA}
\author{Jeremy Rowlette}
\author{H. Ted Stinson}
\author{Michael Pushkarsky}
\author{Timothy Day}
\affiliation{DRS Daylight Solutions, San Diego, CA 92128, USA}
\author{Federico Capasso}
\affiliation{Harvard John A. Paulson School of Engineering and Applied Sciences, Harvard University, Cambridge, MA 02138, USA}
\author{Henry O. Everitt}
\email{capasso@seas.harvard.edu,everitt@phy.duke.edu}
\affiliation{DEVCOM Army Research Lab, Houston, TX  77005, USA}
\affiliation{Department of Physics, Duke University, Durham, NC 27708, USA}





\maketitle
\tableofcontents

This document contains additional tables and figures to support the content of the main text. It contains supporting informations regarding the equipement and the methods used to obtain the results presented in the manuscript.

\section{Experimental details}

The schematic shown in Figure 3 completely describe the experimental setup with the exception of the technique used to set the gas pressure in the cavity. First the gas cell and the laser cavity are brought to a high vacuum ($P<10^{-5}$) Torr with rotary vane pump (Agilent DS102) and a turbo molecular pump (Agilent Turbo V81). Pure anhydrous ammonia is contained in a small 10 mL stainless steel  cylinder connected to the system with a Buna-N diaphragm valve (VAT). This cylinder is cooled down by liquid nitrogen which causes the gas to solidify. After this initial cool down, the valve to the vacuum pump is closed and the valve of the cylinder is opened. Then the liquid nitrogen is removed and the cylinder is allowed to warm, which causes the gas to sublime and fill the rest of the system. The valve of the cylinder is closed when the proper pressure is achieved. The pressure in the cell is measured by a thermocouple Gauge (Alcatel sensor AP1004 and controller ACR1000) and an absolute manometer (MKS Bar-a-tron, range 1 Torr, heated to 40$^\circ$C). To remove gas from the system, the cylinder is cooled again and the needle valve is opened until the gas is recovered and the laser system is empty. 

The pump QCL used in this work was factory modified to enable precise analog tuning of the emission frequency. Experimentally, the QCL frequency was first roughly tuned within 0.1cm$^{-1}$ of the desired infrared transition using the internal grating of the EC-QCL and then was finely tuned by temperature tuning. Temperature tuning resolution of 0.1 K was typically necessary to tune the QCL emission on an IR transition. The precise tuning of the emission frequency was achieved by monitoring the transmission of a small part of the beam (1-5\%) through a gas cell containing 100 mTorr of ammonia.

The Schottky diodes detector used is a Virginia Diodes Inc. ZBD-F WR1.0, with typical characteristics available online at https://www.vadiodes.com/en/products/detectors?id=218. The ZBD signal is amplified by a voltage amplfier with an adjustable gain (SRS 560). For most experiment the typical gain was between 2000 and 20000 with a 30 Hz low-pass filter.

The receiver used in this work is a Virginia Diodes Inc. SAX WR1.0, with typical characteristics available online at https://www.vadiodes.com/index.php/en/products/spectrum-analyzer?id=659. For heterodyne frequency measurements, the local oscillator is provided by a tunable frequency generator (Wiltron 68347B) while the intermediate frequency is measured on a spectrum analyzer (Agilent E4448A). Both instruments are provided with an external 10 MHz frequency reference, from a GPS disciplined oscillator (SRS FS752).

The power meter used to measure the laser output power is a Virginia Diodes Inc. PM5B, with a conical horn antenna matching the WR10 waveguide input of the power meter head.

The TPX window sealing the rear output coupler was made from a 25mm diameter rod of TPX purchased from MilliporeSigma, cut to size on a lathe and then lapped and polished with abrasive paper.

\section{Beam profile}

\begin{figure}[ht]
\centering
\includegraphics[width=0.5\linewidth]{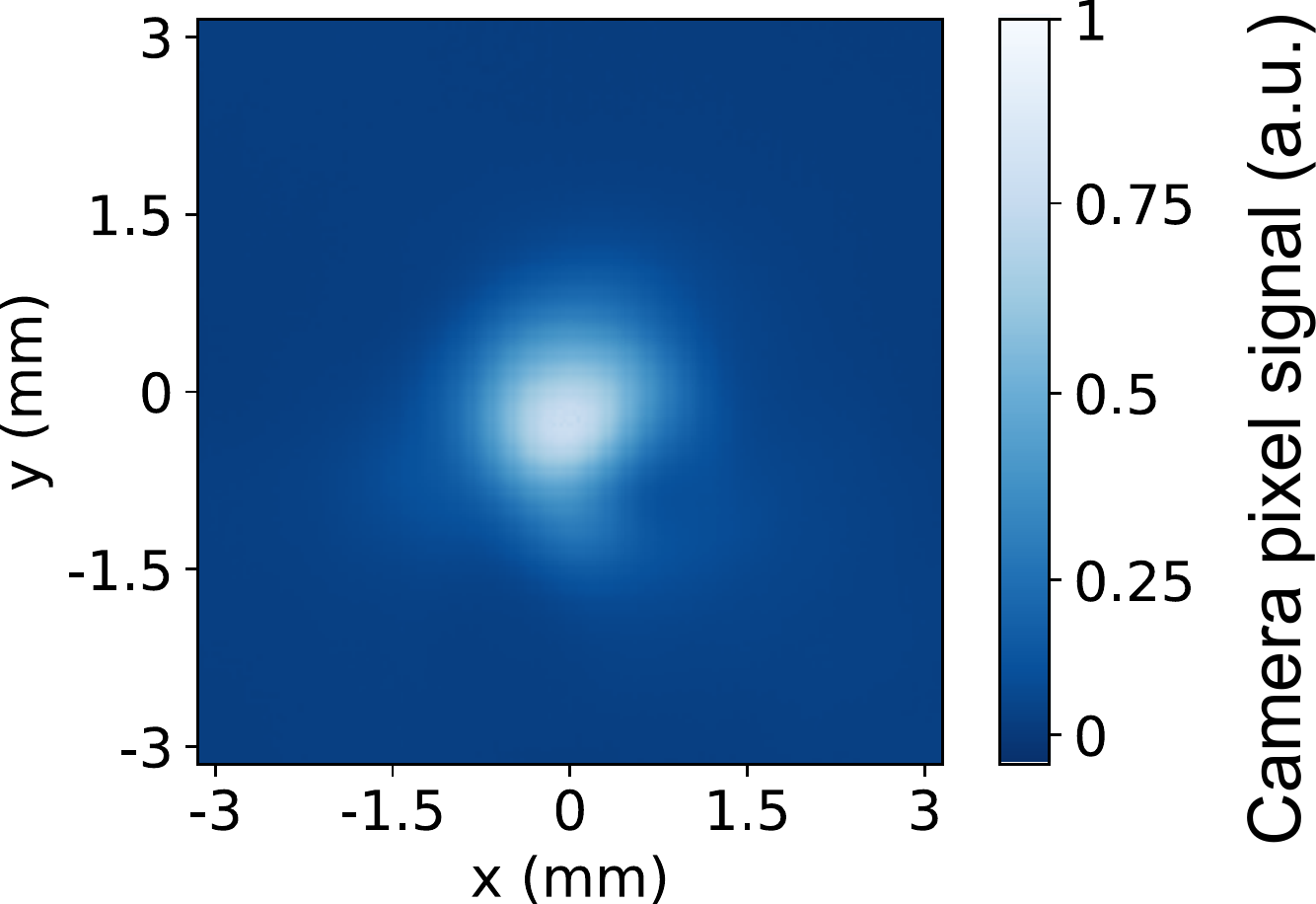}
\caption{Plot of the measured spot profile with a THz camera after focusing light with a 150 mm focal length lens. This profile can be fitted by a Gaussian spot profile with radii 630 \textmu m, and 875 \textmu m.}
\label{fig:figSbeam}
\end{figure}

The profile of a focused THz spot at 3.37 THz has been measured by placing a 50 mm diameter, 150 mm focal length TPX lens, in between the pinhole coupler and a THz sensitive camera (INO Microxcam 384i-THz with 35 \textmu m pixel pitch). The distance between the lens and the camera and pinhole was set at 300 mm (twice the focal length), thereby realizing a 4f system.
The measured beam profile is shown in Fig~\ref{fig:figSbeam}.

\section{Laser transition frequencies and line-by-line performance}

\begin{figure*}[ht]
\centering
\includegraphics[width=\linewidth]{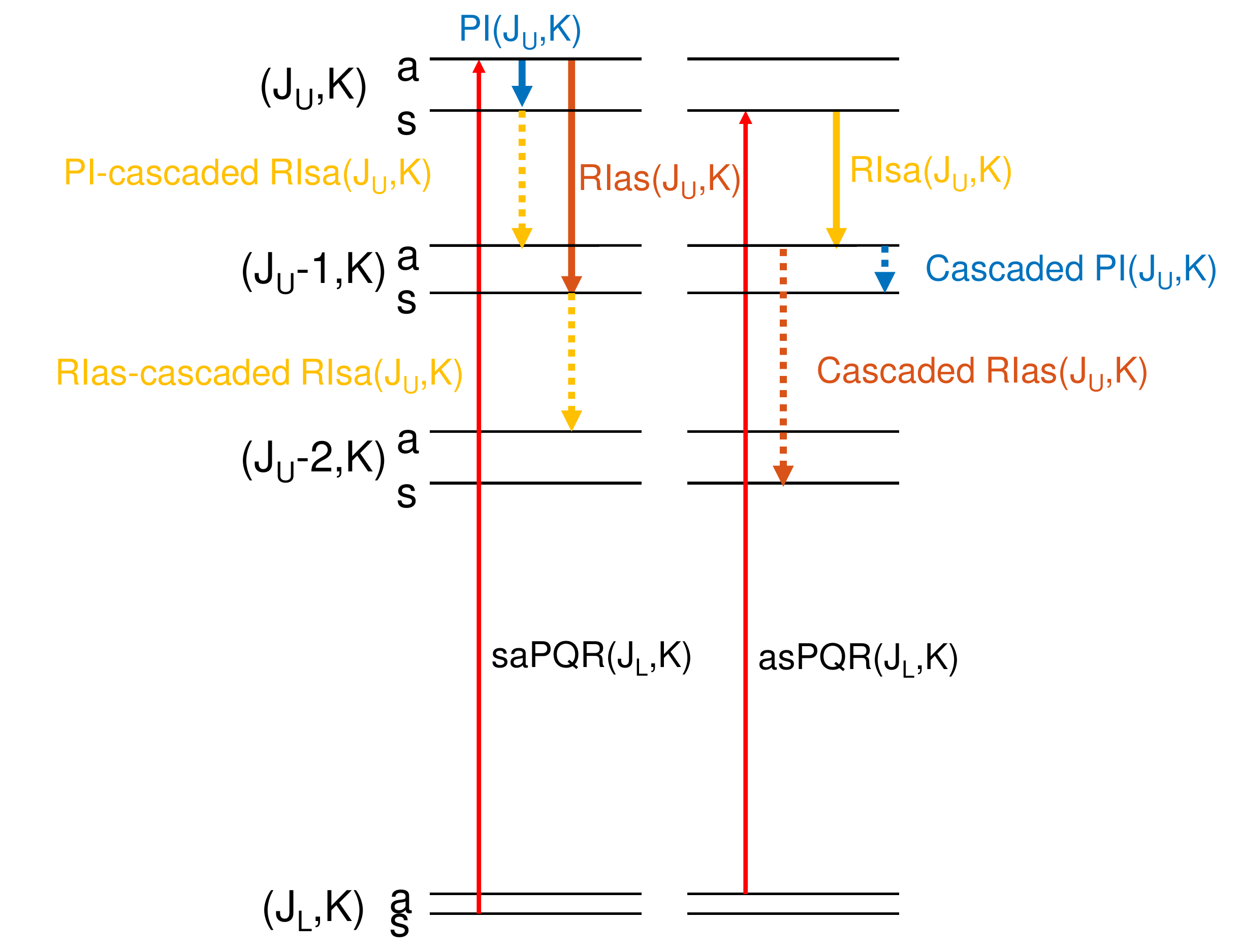}
\caption{Three types of laser transitions are possible in NH$_3$ QPMLs: pure inversion (PI, blue) and rotation-inversion (RIas, red) when an s $\to$ a IR transition is pumped, and rotation-inversion (RIsa, yellow) for an a $\to$ s IR pump.  Cascaded transitions for each type are illustrated as dashed arrows of the same colors. When an sa pump is used, two types of cascaded transitions can occur: a PI-cascaded RIsa and an asRI-cascaded RIsa. When an as pump is used two types of cascaded transitions can occur: a cascaded PI transition or a cascaded RIas transition.}
\label{fig:figS0}
\end{figure*}

Figure.\ref{fig:figS0} recalls the labels for direct and cascaded lasing transition in ammonia QPMLs.

The method to obtain the lasing frequencies, normalized threshold and normalized slope efficiency from the rotational constants is given in Ref.~\citenum{amirzhan2022quantum} and is recalled and adapted for the ammonia molecule in the following.

\subsection{Simple model}

First, we recall the equation of a simple model that was derived in Ref.\citenum{chevalier2019widely} in Eq.\ref{eq:pthz}, that gives the THz emission power $P_{THz}$ at the emission frequency $\nu_{THz}$:

\begin{equation}
P_{THz}=    \frac{T}{4} \frac{ \nu_{THz}}{\nu_{IR}} \frac{ \alpha_{IR}}{\alpha_{cell} } \left (P_{QCL} - P_{th} \right )
    \label{eq:pthz}
\end{equation}
where $\alpha_{IR}$ is the IR absorption of the gain medium at the pump frequency $\nu_{IR}$, $\alpha_{\mathrm{cell}}$ represents the total THz losses of the laser cavity, $T$ is the output coupling fraction, $P_{QCL}$ is the QCL pump power absorbed by the gain medium (which corresponds to the injected IR power minus the power absorbed by THz cavity walls), and $P_{th}$ is the threshold power as expressed below:
\begin{equation}
P_{th}=    \frac{h}{4 \pi} \frac{ \nu_{IR}}{\alpha_{IR}} (\alpha_{\mathrm{cell}} R_{\mathrm{cell}}) \frac{ u^2}{ | \langle J_L |  \mu |  J_U \rangle |^2} 
    \label{eq:pth}
\end{equation}
where $R_{\mathrm{cell}} $ is the radius of the THz cavity, $u$ is the average absolute molecular velocity, and $\langle J_L |  \mu |  J_U \rangle $ is the transition dipole matrix element of the lasing rotational transition.

From Eq.~\ref{eq:pthz}, one can define the total power efficiency $\eta$, as:

\begin{equation}
\eta= \frac{T}{4} \frac{ \nu_{THz}}{\nu_{IR}} \frac{ \alpha_{IR}}{\alpha_{cell} }
    \label{eq:eta}
\end{equation}

The lasing transition dipole element can be expressed as a function of the molecular permanent dipole moment $\mu_0$, the inversion symmetries $\sigma$ and $\sigma'$, and the $J$ and $K$ quantum numbers~\cite{townes1975microwave}, in Eq.~\ref{eq:dipoleLasingRI} for rotation-inversion (RI) lines and in Eq.~\ref{eq:dipoleLasingPI} for pure inversion (PI) lines.

\begin{equation}
  | \langle \sigma,J,K,V| \mu | \sigma',J+1, K, V \rangle |_{RI} ^ 2 = \mu_0^2 \frac{(J+1)^2 - K^2}{(J+1)(2J+3)}, \sigma \neq \sigma' 
    \label{eq:dipoleLasingRI}
\end{equation}

\begin{equation}
  | \langle \sigma,J,K,V| \mu | \sigma',J+1, K, V \rangle |_{PI} ^ 2 = \mu_0^2 \frac{K^2}{J(J+1)}
    \label{eq:dipoleLasingPI}
\end{equation}

One should note that a enhanced model~\cite{wang2021optimizing} that incorporates effects of pump saturation, dipole-dipole collisions and mutiple passes of the infrared the pump in the laser cell, can be used to estimate the THz emission with a better accuracy. However the simplicity of the model given in Eq.\ref{eq:pthz} and Eq.\ref{eq:pth}, makes it suited to consider the QPML performance of different lines in the same laser cavity.

\subsection{Strength of the infrared transitions}

If one assumes that the cavity losses and coupler transmission are frequency independent, the simple model given in equation~\ref{eq:pthz} allows to compare different transitions by comparing the factor $\frac{\nu_{THz}}{\nu_{IR}}  \alpha_{IR}$ . Under the same assumption, the thresholds of different transitions can be compared by comparing the factor $\frac{ \nu_{IR}}{\alpha_{IR} | \langle J_L |  \mu |  J_U \rangle |^2 }$

The frequencies $\nu_{THz}$ and $\nu_{IR}$ can be found in published tables of transitions, so the remaining challenge is to compute a relative infrared absorption $\alpha_{IR}$, which can be calculated as follows:

\begin{equation}
\alpha_{IR} =    \frac{1-\exp(-\alpha_0 L) }{L}
    \label{eq:eqAlpha}
\end{equation}
where $L$ is the cavity length, and $\alpha_0$ is the IR absorption coefficient for a given ro-vibrational transition and is proportional to the square of the transition dipole matrix element~\cite{atkins2011molecular}:

\begin{equation}
\alpha_{0}((\sigma,J,K,V) \rightarrow (\sigma',J',K',V'))  \propto | \langle \sigma,J,K,V| \mu | \sigma',J', K', V'\rangle | ^ 2 
    \label{eq:eq5}
\end{equation}

In the case where either the cell is short or the absorption strength is small, we have $\alpha_0 L\ll 1$, which leads to $\alpha_{IR} \approx \alpha_0 $.
The absorption coefficient $\alpha_0$ strongly depends on the population of energy levels, which is determined by the Boltzmann distribution and the molecular degeneracy.

The value of the dipole matrix element of an IR ro-vibrational transition depends on the type of the transition (P, Q, or R) the initial and final inversion symmetry states $\sigma$, and the initial and final quantum numbers~\cite{townes1975microwave} : 

\begin{equation}
  | \langle \sigma,J-1,K,V+1| \mu | \sigma',J, K, V \rangle | ^ 2_P \propto \mu_0^2  \frac{J^2-K^2}{J(2J+1)} \quad \sigma \neq \sigma'
    \label{eq:dipoleP}
\end{equation}

\begin{equation}
  | \langle \sigma,J,K,V+1| \mu | \sigma',J, K, V \rangle | ^ 2_Q \propto \mu_0^2 \frac{K^2}{J(J+1)} \quad  \sigma \neq \sigma'
    \label{eq:dipoleQ}
\end{equation}

\begin{equation}
  | \langle \sigma,J+1,K,V+1| \mu | \sigma',J, K, V \rangle | ^ 2_R \propto \mu_0^2 \frac{(J+1)^2 - K^2}{(J+1)(2J+1)} \quad \sigma \neq \sigma'
    \label{eq:dipoleR}
\end{equation}

The relative IR absorption line strength used to estimate the relative QPML performance for a given $\sigma,J,K$ is estimated by the following expression:

\begin{equation}
  L(\sigma_{initial},J_{initial},\sigma_{final},J_{final},K)=  | \langle \sigma_{final},J_{final},K,V+1| \mu | \sigma_{initial},J_{initial}, K, V \rangle | ^ 2 Pop(\sigma_{initial},J_{initial},K),
    \label{eq:linestrength}
\end{equation}

where $Pop(\sigma,J,K)$ is the the relative thermal population of the ground state energy level with quantum numbers J and K and symmetry state $\sigma$. This population factor is expressed similarly as in Ref.~\citenum{amirzhan2022quantum}, as a function of the ground state energy levels for instance using the rotational constants from the ammonia molecule\cite{yu2010submillimeter}, the correct energy level degeneracy, the gas temperature $T$ and the Boltzmann constant $k_B$.

\subsection{Normalized power efficiency and threshold}

The normalized power efficiency is then calculated as follows for the R-branch pumped PI, asRI, and the saRI lasing lines, respectively based on formulas derived in Ref.\citenum{amirzhan2022quantum}: the relative IR absorption line strength is multiplied by the relative lasing frequency to pump frequency ratio as a function of the pumped quantum number $J_{L}$ and $K$: 

\begin{equation}
\begin{split}
  \eta_{PI}(J_{L},K) &= \frac{\nu_{PI}(J_{L}+1,K)}{\nu_{saR}(J_L,K)} L(s,J_{L},a,J_{L}+1,K) \\
  &= \frac{\nu_{PI}(J_{L}+1,K)}{\nu_{saR}(J_L,K)} | \langle a,J_{L}+1,K,V+1| \mu | s,J_{L}, K, V \rangle | ^ 2 Pop(s,J_{L},K)
    \label{eq:RbranchPE_PI}
    \end{split}
\end{equation}

\begin{equation}
\begin{split}
  \eta_{asRI}(J_{L},K) &= \frac{\nu_{asRI}(J_{L}+1,K)}{\nu_{saR}(J_L,K)} L(s,J_{L},a,J_{L}+1,K) \\
  &= \frac{\nu_{asRI}(J_{L}+1,K)}{\nu_{saR}(J_L,K)} | \langle a,J_{L}+1,K,V+1| \mu | s,J_{L}, K, V \rangle | ^ 2 Pop(s,J_{L},K)
    \label{eq:RbranchPE_asRI}
    \end{split}
\end{equation}

\begin{equation}
\begin{split}
  \eta_{saRI}(J_{L},K) &= \frac{\nu_{saRI}(J_{L}+1,K)}{\nu_{asR}(J_L,K)} L(a,J_{L},s,J_{L}+1,K) \\
  &= \frac{\nu_{saRI}(J_{L}+1,K)}{\nu_{asR}(J_L,K)} | \langle s,J_{L}+1,K,V+1| \mu | a,J_{L}, K, V \rangle | ^ 2 Pop(a,J_{L},K)
    \label{eq:RbranchPE_saRI}
    \end{split}
\end{equation}

The normalized lasing threshold is calculated for R branch pumped asRI, saRI and PI transitions, by taking the ratio of the infrared pump frequency by the line strength and the lasing matrix element as a function of the pumped quantum numbers $J_{L}$ and $K$. Q-branch pumped transitions are computed in a similar way. 

\begin{equation}
\begin{split}
  Th_{PI}(J_{L},K) &= \frac{\nu_{saR}(J_L,K)} { | \langle s,J_L,K,V| \mu | a,J_L, K, V \rangle | ^ 2 L(s,J_{L},a,J_{L}+1,K) } \\ 
  &=  \frac{\nu_{saR}(J_L,K)} { | \langle s,J_L,K,V| \mu | a,J_L, K, V \rangle | ^ 2  | \langle a,J_{L}+1,K,V+1| \mu | s,J_{L}, K, V \rangle | ^ 2 Pop(s,J_{L},K)}
    \label{eq:Rbranchth_PI}
    \end{split}
\end{equation}

\begin{equation}
\begin{split}
  Th_{asRI}(J_{L},K) &= \frac{\nu_{saR}(J_L,K)} { | \langle s,J_L,K,V| \mu | a,J_L+1, K, V \rangle | ^ 2 L(s,J_{L},a,J_{L}+1,K) } \\ 
  &=  \frac{\nu_{saR}(J_L,K)} { | \langle s,J_L,K,V| \mu | a,J_L+1, K, V \rangle | ^ 2  | \langle a,J_{L}+1,K,V+1| \mu | s,J_{L}, K, V \rangle | ^ 2 Pop(s,J_{L},K)}
    \label{eq:Rbranchth_asRI}
    \end{split}
\end{equation}

\begin{equation}
\begin{split}
  Th_{saRI}(J_{L},K) &= \frac{\nu_{asR}(J_L,K)} { | \langle a,J_L,K,V| \mu | s,J_L+1, K, V \rangle | ^ 2 L(a,J_{L},s,J_{L}+1,K) } \\ 
  &=  \frac{\nu_{asR}(J_L,K)} { | \langle a,J_L,K,V| \mu | s,J_L+1, K, V \rangle | ^ 2  | \langle s,J_{L}+1,K,V+1| \mu | a,J_{L}, K, V \rangle | ^ 2 Pop(a,J_{L},K)}
    \label{eq:Rbranchth_asRI}
    \end{split}
\end{equation}

\subsection{Plots of line frequencies and relative strength}

\begin{figure*}[ht]
\centering
\includegraphics[width=\linewidth]{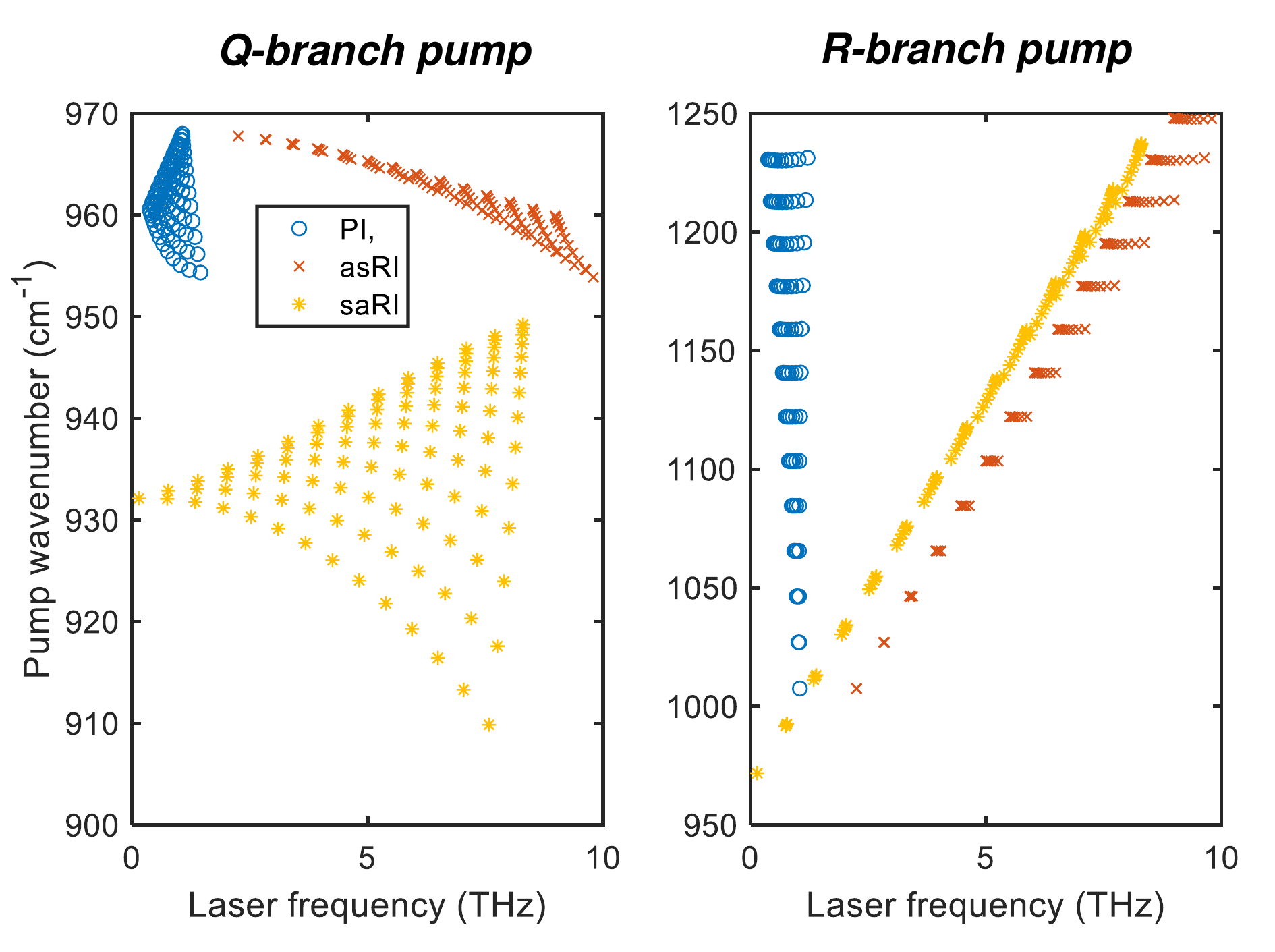}
\caption{Direct lasing transition frequencies for the NH$_3$ QPML a Q-branch or an R-branch pump.}
\label{fig:figS1}
\end{figure*}

\begin{figure*}[ht]
\centering
\includegraphics[width=\linewidth]{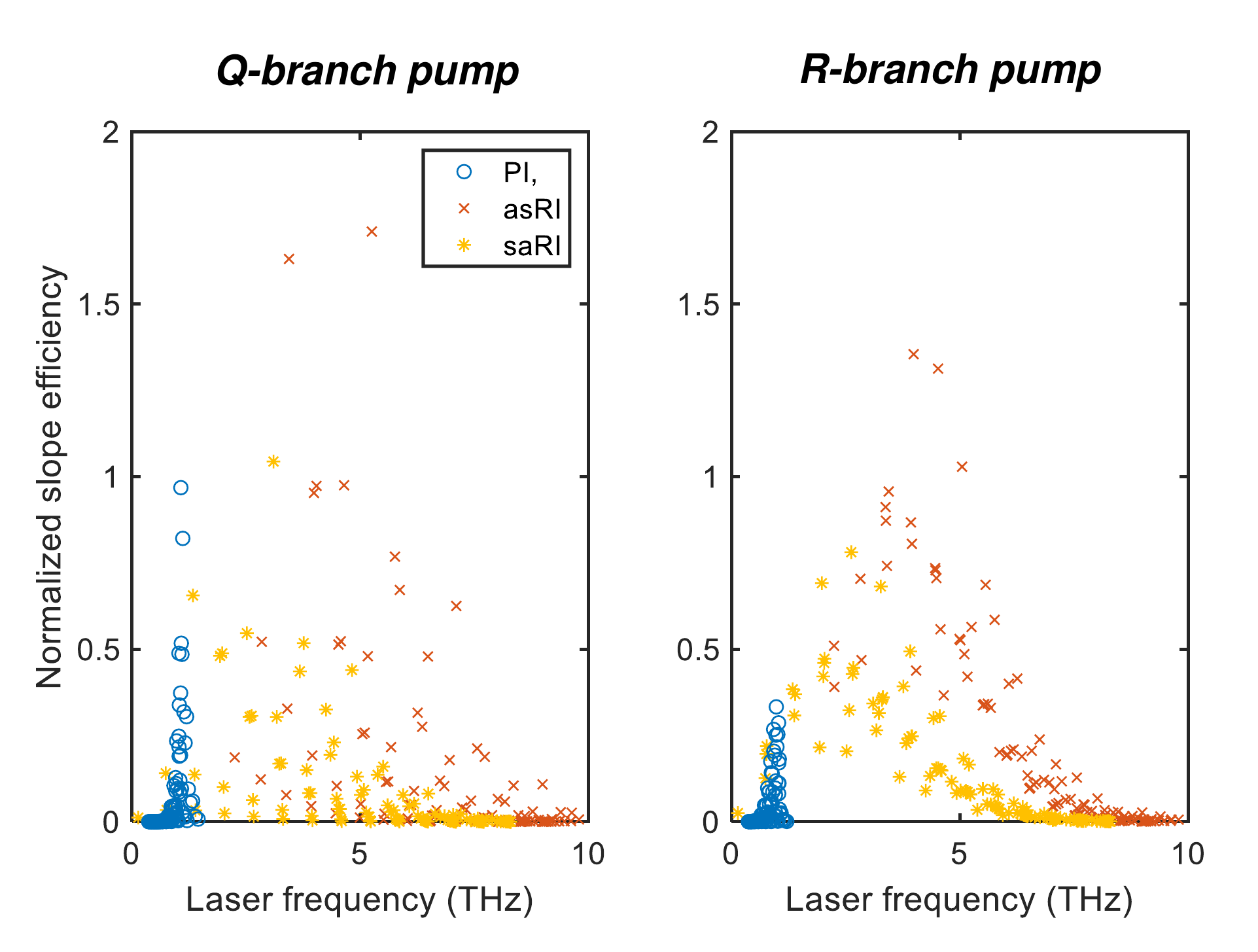}
\caption{Normalized slope efficiency  of the NH$_3$ QPML as function of the transition frequency, using the method of Ref.~\citenum{amirzhan2022quantum} for a Q-branch or an R-branch pump. }
\label{fig:figS2}
\end{figure*}

Figures~\ref{fig:figS1}-\ref{fig:figS2}  along with the numbers in Tables~\ref{tab:tabsaR}-\ref{tab:tabasQ} were obtained using published transitions frequencies from the HITRAN database\cite{HITRAN2020} and the JPL database\cite{Splatalogue2007}. These sources include measurements from other sources\cite{pearson2016modeling,yu2010submillimeter}

\begin{figure*}[ht]
\centering
\includegraphics[width=\linewidth]{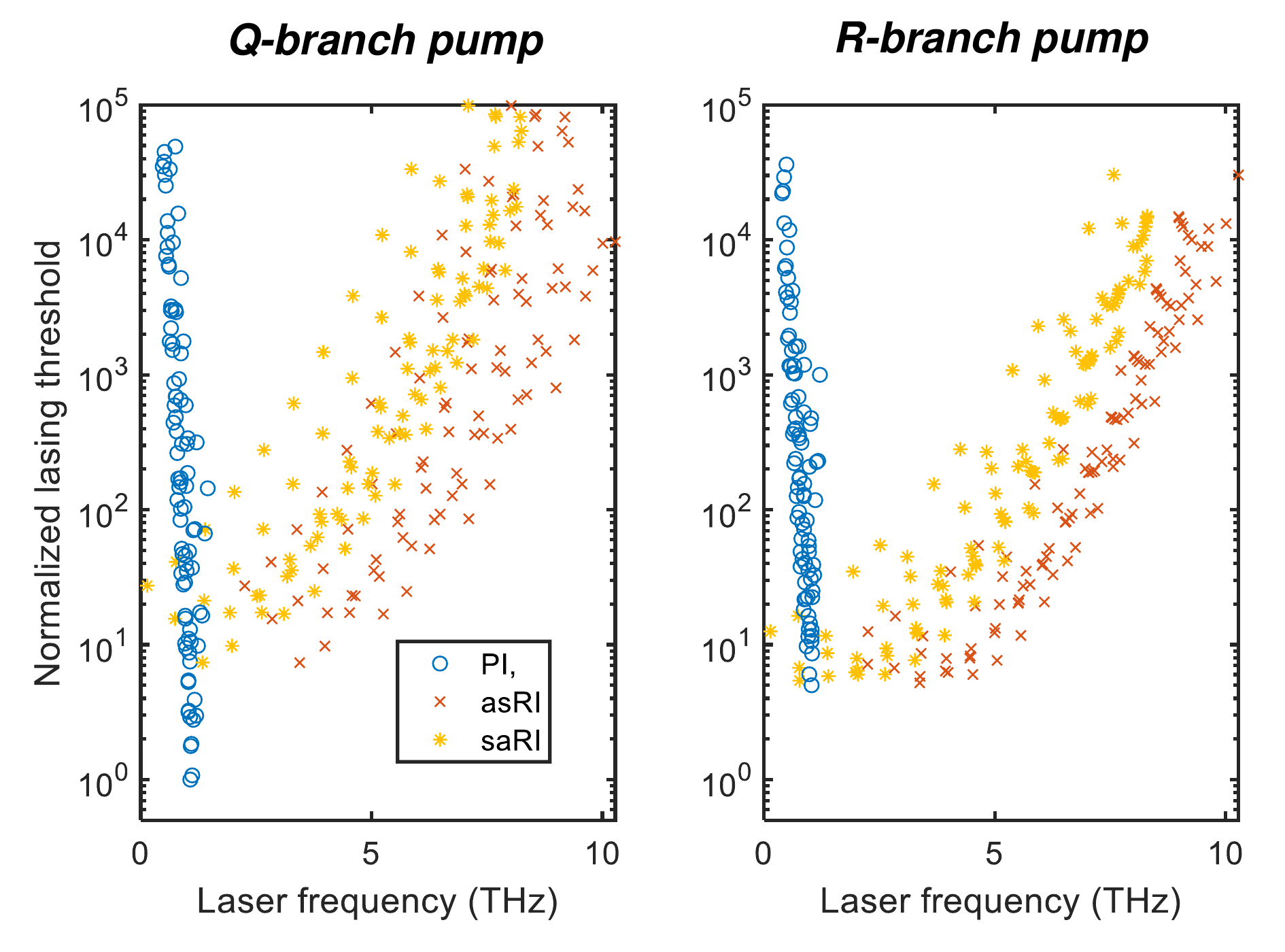}
\caption{Normalized lasing threshold  of the NH$_3$ QPML as function of the transition frequency, using the method of Ref.~\citenum{amirzhan2022quantum} for a Q-branch or an R-branch pump.}
\label{fig:fig3}
\end{figure*}

\begingroup
\squeezetable
\begin{table*}[htp]
\caption{\label{tab:tabsaR} Frequencies of RIas and PI direct lasing and RIsa cascaded lasing, as a function of the saR infrared transitions wavenumber. The $\dagger$ symbol indicates pump transitions that were successfully pumped (i.e. a laser line was obtained). The $\ddagger$ symbol indicates the lines that were presumed to be have been lasing owing to the selection rules and the bandwidth of the detector. The $\mathsection$ (resp. $\mathparagraph$) symbol highlights lines that were made to lase experimentally and for which the emission has been confirmed using the receiver (resp. the scanning Fabry Perot interferometer).}
\begin{tabular*}{0.70\textwidth}{c|c|c|c|c|c|c|c}
\hline\hline
saR pump & $J_L$ & $J_U$ & $K$ & Lasing frequency & Lasing frequency & Lasing frequency  & Lasing frequency  \\
wavenumber (cm$^{-1}$) & & &  & direct PI (THz) & direct RIas (THz)  & PI-cascaded RIsa (THz) & RIas-cascaded RIsa (THz)\\
\hline
1007.547$^\dagger$ & 1  & 2  & 0  & -        & 2.244466$^\ddagger$ & -        & -        \\
1007.540$^\dagger$ & 1  & 2  & 1  & 1.045319$^\mathsection$ & 2.252112 & 0.140142 & -        \\
1027.047$^\dagger$ & 2  & 3  & 1  & 1.014085$^\mathsection$ & 2.822258 & 0.762852 & 0.140142 \\
1027.033$^\dagger$ & 2  & 3  & 2  & 1.035816$^\mathsection$ & 2.845290 & 0.741788 & -        \\
1046.405$^\dagger$ & 3  & 4  & 0  & -        & 3.373613$^\mathparagraph$ & -        & 0.769710$^\mathsection$ \\
1046.401$^\dagger$ & 3  & 4  & 1  & 0.973827$^\mathsection$ & 3.380990$^\ddagger$ & 1.393079 & 0.762852$^\mathsection$ \\
1046.388$^\dagger$ & 3  & 4  & 2  & 0.994748$^\mathsection$ & 3.403613 & 1.373049 & 0.741788 \\
1046.375$^\dagger$ & 3  & 4  & 3  & 1.030530$^\mathsection$ & 3.442259 & 1.338679 & -        \\
1065.594$^\dagger$ & 4  & 5  & 1  & 0.925658$^\dagger$ & 3.928686$^\mathparagraph$ & 2.029201 & 1.393079$^\mathparagraph$ \\
1065.582$^\dagger$ & 4  & 5  & 2  & 0.945605$^\mathsection$ & 3.950724 & 2.010371 & 1.373049 \\
1065.565$^\dagger$ & 4  & 5  & 3  & 0.979650$^\mathsection$ & 3.988292$^\ddagger$ & 1.978112 & 1.338679 \\
1065.564$^\dagger$ & 4  & 5  & 4  & 1.029375$^\mathsection$ & 4.043027 & 1.931049 & -        \\
1084.629$^\dagger$ & 5  & 6  & 0  & -        & 4.458880$^\ddagger$ & -        & 2.035454$^\ddagger$ \\
1084.624 & 5  & 6  & 1  & 0.870878 & 4.465963 & 2.669411 & 2.029201 \\
1084.610 & 5  & 6  & 2  & 0.889711 & 4.487253 & 2.651944 & 2.010371 \\
1084.593 & 5  & 6  & 3  & 0.921941 & 4.523619 & 2.622003 & 1.978112 \\
1084.584 & 5  & 6  & 4  & 0.968809 & 4.576377 & 2.578131 & 1.931049 \\
1084.599 & 5  & 6  & 5  & 1.032322 & 4.647576 & 2.518617 & -        \\
1103.486 & 6  & 7  & 1  & 0.810919 & 4.993576 & 3.311716 & 2.669411 \\
1103.470 & 6  & 7  & 2  & 0.828525 & 5.013987 & 3.295760 & 2.651944 \\
1103.441 & 6  & 7  & 3  & 0.858394 & 5.048600 & 3.268236 & 2.622003 \\
1103.430 & 6  & 7  & 4  & 0.902459 & 5.099335 & 3.228040 & 2.578131 \\
1103.434 & 6  & 7  & 5  & 0.961885 & 5.167540 & 3.173290 & 2.518617 \\
1103.480 & 6  & 7  & 6  & 1.039360 & 5.256011 & 3.101558 & -        \\
1122.185 & 7  & 8  & 0  & -        & 5.505906 & -        & 3.317199 \\
1122.178 & 7  & 8  & 1  & 0.747287 & 5.512472 & 3.954265 & 3.311716 \\
1122.160 & 7  & 8  & 2  & 0.763583 & 5.531898 & 3.939779 & 3.295760 \\
1122.133 & 7  & 8  & 3  & 0.791531 & 5.565008 & 3.915058 & 3.268236 \\
1122.104 & 7  & 8  & 4  & 0.831969 & 5.612952 & 3.878471 & 3.228040 \\
1122.094 & 7  & 8  & 5  & 0.887018 & 5.677725 & 3.828821 & 3.173290 \\
1122.118 & 7  & 8  & 6  & 0.958828 & 5.761786 & 3.763547 & 3.101558 \\
1122.204 & 7  & 8  & 7  & 1.050521 & 5.868273 & 3.679575 & -        \\
1140.700 & 8  & 9  & 1  & 0.681503 & 6.023669 & 4.594879 & 3.954265 \\
1140.679 & 8  & 9  & 2  & 0.696434 & 6.042017 & 4.581997 & 3.939779 \\
1140.627 & 8  & 9  & 3  & 0.721263 & 6.072684 & 4.559893 & 3.915058 \\
1140.604 & 8  & 9  & 4  & 0.759001 & 6.118368 & 4.527384 & 3.878471 \\
1140.578 & 8  & 9  & 5  & 0.809481 & 6.179461 & 4.482978 & 3.828821 \\
1140.579 & 8  & 9  & 6  & 0.875368 & 6.258671 & 4.424421 & 3.763547 \\
1140.632 & 8  & 9  & 7  & 0.959566 & 6.359053 & 4.348961 & 3.679575 \\
1140.771 & 8  & 9  & 8  & 1.065868 & 6.484720 & 4.252680 & -        \\
1159.056 & 9  & 10 & 0  & -        & 6.522154 & -        & 4.599511 \\
1159.047 & 9  & 10 & 1  & 0.615046 & 6.528189 & 5.231649 & 4.594879 \\
1159.026 & 9  & 10 & 2  & 0.628587 & 6.545438 & 5.220411 & 4.581997 \\
1158.985 & 9  & 10 & 3  & 0.651988 & 6.574681 & 5.201418 & 4.559893 \\
1158.929 & 9  & 10 & 4  & 0.685168 & 6.616804 & 5.172636 & 4.527384 \\
1158.888 & 9  & 10 & 5  & 0.730984 & 6.674009 & 5.133494 & 4.482978 \\
1158.865 & 9  & 10 & 6  & 0.790818 & 6.748090 & 5.081852 & 4.424421 \\
1158.885 & 9  & 10 & 7  & 0.867343 & 6.841979 & 5.015021 & 4.348961 \\
1158.976 & 9  & 10 & 8  & 0.964060 & 6.959425 & 4.929483 & 4.252680 \\
1159.180 & 9  & 10 & 9  & 1.085499 & 7.105332 & 4.820644 & -        \\
1177.218 & 10 & 11 & 1  & 0.549305 & 7.027065 & 5.862714 & 5.231649 \\
1177.198 & 10 & 11 & 2  & 0.561458 & 7.043165 & 5.853142 & 5.220411 \\
1177.111 & 10 & 11 & 3  & 0.580477 & 7.069147 & 5.836679 & 5.201418 \\
1177.078 & 10 & 11 & 4  & 0.611976 & 7.109331 & 5.812187 & 5.172636 \\
1177.021 & 10 & 11 & 5  & 0.653127 & 7.162500 & 5.778388 & 5.133494 \\
1176.977 & 10 & 11 & 6  & 0.706898 & 7.231358 & 5.733593 & 5.081852 \\
1176.964 & 10 & 11 & 7  & 0.775727 & 7.318466 & 5.675430 & 5.015021 \\
1177.008 & 10 & 11 & 8  & 0.862810 & 7.427521 & 5.600634 & 4.929483 \\
1177.148 & 10 & 11 & 9  & 0.972301 & 7.562901 & 5.505064 & 4.820644 \\
1177.431 & 10 & 11 & 10 & 1.109546 & 7.730372 & 5.383323 & -       \\
\hline
\end{tabular*}
\end{table*}
\endgroup

\begingroup
\squeezetable
\begin{table*}[htp]
\caption{\label{tab:tabasR} Frequencies of RIsa direct lasing and PI and RIas cascaded lasing, as a function of the asR infrared transitions wavenumber. The $\dagger$ symbol indicates pump transitions that were successfully pumped (i.e. a laser line was obtained). The $\ddagger$ symbol indicates the lines that were presumed to be have been lasing owing to the selection rules and the bandwidth of the detector. The $\mathsection$ (resp. $\mathparagraph$) symbol highlights lines that were made to lase experimentally and for which the emission has been confirmed using the receiver (resp. the scanning Fabry Perot interferometer).}
\begin{tabular*}{0.55\textwidth}{c|c|c|c|c|c|c}
\hline\hline
asR pump & $J_L$ & $J_U$ & $K$ & Lasing frequency & Lasing frequency & Lasing frequency  \\
wavenumber (cm$^{-1}$) & & &  & direct RIsa (THz) & cascaded PI (THz)  & cascaded RIas (THz)\\
\hline
971.882  & 1  & 2  & 1  & 0.140142 & 1.066651 & -        \\
992.699$^\dagger$  & 2  & 3  & 0  & 0.769710 $^\mathsection$ & -        & 2.244466 \\
992.450  & 2  & 3  & 1  & 0.762852 & 1.045319 & 2.252112 \\
991.691  & 2  & 3  & 2  & 0.741788 & 1.067677 & 0.000000 \\
1013.176 & 3  & 4  & 1  & 1.393079 & 1.014085 & 2.822258 \\
1012.445$^\dagger$ & 3  & 4  & 2  & 1.373049$^\ddagger$ & 1.035816 & 2.845290 \\
1011.204$^\dagger$ & 3  & 4  & 3  & 1.338679$^\ddagger$ & 1.073050$^\mathsection$ & -        \\
1034.245$^\dagger$ & 4  & 5  & 0  & 2.035454$^\ddagger$ & -        & 3.373613 \\
1034.013 & 4  & 5  & 1  & 2.029201 & 0.973827 & 3.380990 \\
1033.316$^\dagger$ & 4  & 5  & 2  & 2.010371$^\ddagger$ & 0.994748 & 3.403613 \\
1032.131$^\dagger$ & 4  & 5  & 3  & 1.978112$^\ddagger$ & 1.030530$^\mathsection$ & 3.442259 \\
1030.422 & 4  & 5  & 4  & 1.931049 & 1.082592 & -        \\
1054.913 & 5  & 6  & 1  & 2.669411 & 0.925658 & 3.928686 \\
1054.253 & 5  & 6  & 2  & 2.651944 & 0.945605 & 3.950724 \\
1053.130$^\dagger$ & 5  & 6  & 3  & 2.622003$^\ddagger$ & 0.979650$^\mathsection$ & 3.988292 \\
1051.512 & 5  & 6  & 4  & 2.578131 & 1.029375 & 4.043027 \\
1049.346 & 5  & 6  & 5  & 2.518617 & 1.096590 & -        \\
1076.033 & 6  & 7  & 0  & 3.317199 & -        & 4.458880 \\
1075.823 & 6  & 7  & 1  & 3.311716 & 0.870878 & 4.465963 \\
1075.203 & 6  & 7  & 2  & 3.295760 & 0.889711 & 4.487253 \\
1074.149$^\dagger$ & 6  & 7  & 3  & 3.268236$^\ddagger$ & 0.921941$^\mathsection$ & 4.523619 \\
1072.627 & 6  & 7  & 4  & 3.228040 & 0.968809 & 4.576377 \\
1070.591 & 6  & 7  & 5  & 3.173290 & 1.032322 & 4.647576 \\
1067.974 & 6  & 7  & 6  & 3.101558 & 1.115081 & -        \\
1096.690 & 7  & 8  & 1  & 3.954265 & 0.810919 & 4.993576 \\
1096.113 & 7  & 8  & 2  & 3.939779 & 0.828525 & 5.013987 \\
1095.129 & 7  & 8  & 3  & 3.915058 & 0.858394 & 5.048600 \\
1093.711 & 7  & 8  & 4  & 3.878471 & 0.902459 & 5.099335 \\
1091.812 & 7  & 8  & 5  & 3.828821 & 0.961885 & 5.167540 \\
1089.370 & 7  & 8  & 6  & 3.763547 & 1.039360 & 5.256011 \\
1086.304 & 7  & 8  & 7  & 3.679575 & 1.138211 & -        \\
1117.648 & 8  & 9  & 0  & 4.599511 & -        & 5.505906 \\
1117.459 & 8  & 9  & 1  & 4.594879 & 0.747287 & 5.512472 \\
1116.927 & 8  & 9  & 2  & 4.581997 & 0.763583 & 5.531898 \\
1116.020 & 8  & 9  & 3  & 4.559893 & 0.791531 & 5.565008 \\
1114.707 & 8  & 9  & 4  & 4.527384 & 0.831969 & 5.612952 \\
1112.949 & 8  & 9  & 5  & 4.482978 & 0.887018 & 5.677725 \\
1110.689 & 8  & 9  & 6  & 4.424421 & 0.958828 & 5.761786 \\
1107.849 & 8  & 9  & 7  & 4.348961 & 1.050521 & 5.868273 \\
1104.333 & 8  & 9  & 8  & 4.252680 & 1.166164 & -        \\
1138.078 & 9  & 10 & 1  & 5.231649 & 0.681503 & 6.023669 \\
1137.592 & 9  & 10 & 2  & 5.220411 & 0.696434 & 6.042017 \\
1136.757 & 9  & 10 & 3  & 5.201418 & 0.721263 & 6.072684 \\
1135.556 & 9  & 10 & 4  & 5.172636 & 0.759001 & 6.118368 \\
1133.944 & 9  & 10 & 5  & 5.133494 & 0.809481 & 6.179461 \\
1131.870 & 9  & 10 & 6  & 5.081852 & 0.875368 & 6.258671 \\
1129.262 & 9  & 10 & 7  & 5.015021 & 0.959566 & 6.359053 \\
1126.029 & 9  & 10 & 8  & 4.929483 & 1.065868 & 6.484720 \\
1122.055 & 9  & 10 & 9  & 4.820644 & 1.199163 & -        \\
1158.667 & 10 & 11 & 0  & 5.866699 & -        & 6.522154 \\
1158.494 & 10 & 11 & 1  & 5.862714 & 0.615046 & 6.528189 \\
1158.058 & 10 & 11 & 2  & 5.853142 & 0.628587 & 6.545438 \\
1157.305 & 10 & 11 & 3  & 5.836679 & 0.651988 & 6.574681 \\
1156.207 & 10 & 11 & 4  & 5.812187 & 0.685168 & 6.616804 \\
1154.741 & 10 & 11 & 5  & 5.778388 & 0.730984 & 6.674009 \\
1152.853 & 10 & 11 & 6  & 5.733593 & 0.790818 & 6.748090 \\
1150.478 & 10 & 11 & 7  & 5.675430 & 0.867343 & 6.841979 \\
1147.533 & 10 & 11 & 8  & 5.600634 & 0.964060 & 6.959425 \\
1143.908 & 10 & 11 & 9  & 5.505064 & 1.085499 & 7.105332 \\
1139.467 & 10 & 11 & 10 & 5.383323 & 1.237466 & -       \\
\hline
\end{tabular*}
\end{table*}
\endgroup

\begingroup
\squeezetable
\begin{table*}[htp]
\caption{\label{tab:tabsaQ} Frequencies of RIas and PI direct lasing and RIsa cascaded lasing, as a function of the saQ infrared transitions wavenumber. The $\dagger$ symbol indicates pump transitions that were successfully pumped (i.e. a laser line was obtained). The $\mathsection$ symbol highlights lines that were made to lase experimentally and for which the emission has been confirmed using the receiver.}
\begin{tabular*}{0.70\textwidth}{c|c|c|c|c|c|c|c}
\hline\hline
saQ pump & $J_L$ & $J_U$ & $K$ & Lasing frequency & Lasing frequency & Lasing frequency & Lasing frequency  \\
wavenumber (cm$^{-1}$) & & &  & direct PI (THz) & direct RIas (THz)  & PI-cascaded RIsa (THz) & RIas-cascaded RIsa (THz)\\
\hline
967.998 & 1  & 1  & 1  & 1.066651 & -        & -        & -        \\
967.775 & 2  & 2  & 1  & 1.045319 & 2.252112 & 0.140142 & -        \\
967.738 & 2  & 2  & 2  & 1.067677 & -        & -        & -        \\
967.449 & 3  & 3  & 1  & 1.014085 & 2.822258 & 0.762852 & 0.140142 \\
967.407$^\dagger$ & 3  & 3  & 2  & 1.035816$^\mathsection$ & 2.845290 & 0.741788 & -        \\
967.346$^\dagger$ & 3  & 3  & 3  & 1.073050$^\mathsection$ & -        & -        & -        \\
967.031 & 4  & 4  & 1  & 0.973827 & 3.380990 & 1.393079 & 0.762852 \\
966.981 & 4  & 4  & 2  & 0.994748 & 3.403613 & 1.373049 & 0.741788 \\
966.905$^\dagger$ & 4  & 4  & 3  & 1.030530$^\mathsection$ & 3.442259 & 1.338679 & -        \\
966.815 & 4  & 4  & 4  & 1.082592 & -        & -        & -        \\
966.532 & 5  & 5  & 1  & 0.925658 & 3.928686 & 2.029201 & 1.393079 \\
966.474 & 5  & 5  & 2  & 0.945605 & 3.950724 & 2.010371 & 1.373049 \\
966.380 & 5  & 5  & 3  & 0.979650 & 3.988292 & 1.978112 & 1.338679 \\
966.269$^\dagger$ & 5  & 5  & 4  & 1.029375$^\mathsection$ & 4.043027 & 1.931049 & -        \\
966.151 & 5  & 5  & 5  & 1.096590 & -        & -        & -        \\
965.968 & 6  & 6  & 1  & 0.870878 & 4.465963 & 2.669411 & 2.029201 \\
965.899 & 6  & 6  & 2  & 0.889711 & 4.487253 & 2.651944 & 2.010371 \\
965.791 & 6  & 6  & 3  & 0.921941 & 4.523619 & 2.622003 & 1.978112 \\
965.652 & 6  & 6  & 4  & 0.968809 & 4.576377 & 2.578131 & 1.931049 \\
965.499 & 6  & 6  & 5  & 1.032322 & 4.647576 & 2.518617 & -        \\
965.354 & 6  & 6  & 6  & 1.115081 & -        & -        & -        \\
965.351 & 7  & 7  & 1  & 0.810919 & 4.993576 & 3.311716 & 2.669411 \\
965.273 & 7  & 7  & 2  & 0.828525 & 5.013987 & 3.295760 & 2.651944 \\
965.138 & 7  & 7  & 3  & 0.858394 & 5.048600 & 3.268236 & 2.622003 \\
964.980 & 7  & 7  & 4  & 0.902459 & 5.099335 & 3.228040 & 2.578131 \\
964.790 & 7  & 7  & 5  & 0.961885 & 5.167540 & 3.173290 & 2.518617 \\
964.596 & 7  & 7  & 6  & 1.039360 & 5.256011 & 3.101558 & -        \\
964.424 & 7  & 7  & 7  & 1.138211 & -        & -        & -        \\
964.699 & 8  & 8  & 1  & 0.747287 & 5.505906 & 3.954265 & 3.311716 \\
964.610 & 8  & 8  & 2  & 0.763583 & 5.512472 & 3.939779 & 3.295760 \\
964.467 & 8  & 8  & 3  & 0.791531 & 5.531898 & 3.915058 & 3.268236 \\
964.268 & 8  & 8  & 4  & 0.831969 & 5.565008 & 3.878471 & 3.228040 \\
964.041 & 8  & 8  & 5  & 0.887018 & 5.612952 & 3.828821 & 3.173290 \\
963.796 & 8  & 8  & 6  & 0.958828 & 5.677725 & 3.763547 & 3.101558 \\
963.559 & 8  & 8  & 7  & 1.050521 & 5.761786 & 3.679575 & -        \\
963.363 & 8  & 8  & 8  & 1.166164 & -        & -        & -        \\
964.023 & 9  & 9  & 1  & 0.681503 & 6.023669 & 4.594879 & 3.954265 \\
963.924 & 9  & 9  & 2  & 0.696434 & 6.042017 & 4.581997 & 3.939779 \\
963.736 & 9  & 9  & 3  & 0.721263 & 6.072684 & 4.559893 & 3.915058 \\
963.534 & 9  & 9  & 4  & 0.759001 & 6.118368 & 4.527384 & 3.878471 \\
963.269 & 9  & 9  & 5  & 0.809481 & 6.179461 & 4.482978 & 3.828821 \\
962.974 & 9  & 9  & 6  & 0.875368 & 6.258671 & 4.424421 & 3.763547 \\
962.670 & 9  & 9  & 7  & 0.959566 & 6.359053 & 4.348961 & 3.679575 \\
962.388 & 9  & 9  & 8  & 1.065868 & 6.484720 & 4.252680 & -        \\
962.171 & 9  & 9  & 9  & 1.199163 & -        & -        & -        \\
963.336 & 10 & 10 & 1  & 0.615046 & 6.528189 & 5.231649 & 4.594879 \\
963.230 & 10 & 10 & 2  & 0.628587 & 6.545438 & 5.220411 & 4.581997 \\
963.057 & 10 & 10 & 3  & 0.651988 & 6.574681 & 5.201418 & 4.559893 \\
962.791 & 10 & 10 & 4  & 0.685168 & 6.616804 & 5.172636 & 4.527384 \\
962.489 & 10 & 10 & 5  & 0.730984 & 6.674009 & 5.133494 & 4.482978 \\
962.144 & 10 & 10 & 6  & 0.790818 & 6.748090 & 5.081852 & 4.424421 \\
961.776 & 10 & 10 & 7  & 0.867343 & 6.841979 & 5.015021 & 4.348961 \\
961.411 & 10 & 10 & 8  & 0.964060 & 6.959425 & 4.929483 & 4.252680 \\
961.086 & 10 & 10 & 9  & 1.085499 & 7.105332 & 4.820644 & -        \\
960.852 & 10 & 10 & 10 & 1.237466 & -        & -        & -       \\
\hline
\end{tabular*}
\end{table*}
\endgroup

\begingroup
\squeezetable
\begin{table*}[htp]
\caption{\label{tab:tabasQ} Frequencies of RIsa direct lasing and PI and RIas cascaded lasing, as a function of the asQ infrared transitions wavenumber}
\begin{tabular*}{0.55\textwidth}{c|c|c|c|c|c|c}
\hline\hline
asQ pump & $J_L$ & $J_U$ & $K$ & Lasing frequency & Lasing frequency & Lasing frequency  \\
wavenumber (cm$^{-1}$) & & &  & direct RIsa (THz) & cascaded PI (THz)  & cascaded RIas (THz)\\
\hline
932.136 & 2  & 2  & 1 & 0.140142 & 1.066651 & -        \\
932.881 & 3  & 3  & 1 & 0.762852 & 1.045319 & 2.252112 \\
932.094 & 3  & 3  & 2 & 0.741788 & 1.067677 & -        \\
933.843 & 4  & 4  & 1 & 1.393079 & 1.014085 & 2.822258 \\
933.076 & 4  & 4  & 2 & 1.373049 & 1.035816 & 2.845290 \\
931.774 & 4  & 4  & 3 & 1.338679 & 1.073050 & -        \\
934.994 & 5  & 5  & 1 & 2.029201 & 0.973827 & 3.380990 \\
934.252 & 5  & 5  & 2 & 2.010371 & 0.994748 & 3.403613 \\
932.992 & 5  & 5  & 3 & 1.978112 & 1.030530 & 3.442259 \\
931.177 & 5  & 5  & 4 & 1.931049 & 1.082592 & -        \\
936.305 & 6  & 6  & 1 & 2.669411 & 0.925658 & 3.928686 \\
935.592 & 6  & 6  & 2 & 2.651944 & 0.945605 & 3.950724 \\
934.380 & 6  & 6  & 3 & 2.622003 & 0.979650 & 3.988292 \\
932.636 & 6  & 6  & 4 & 2.578131 & 1.029375 & 4.043027 \\
930.307 & 6  & 6  & 5 & 2.518617 & 1.096590 & -        \\
937.740 & 7  & 7  & 1 & 3.311716 & 0.870878 & 4.465963 \\
937.059 & 7  & 7  & 2 & 3.295760 & 0.889711 & 4.487253 \\
935.904 & 7  & 7  & 3 & 3.268236 & 0.921941 & 4.523619 \\
934.236 & 7  & 7  & 4 & 3.228040 & 0.968809 & 4.576377 \\
932.011 & 7  & 7  & 5 & 3.173290 & 1.032322 & 4.647576 \\
929.162 & 7  & 7  & 6 & 3.101558 & 1.115081 & -        \\
939.264 & 8  & 8  & 1 & 3.954265 & 0.810919 & 4.993576 \\
938.617 & 8  & 8  & 2 & 3.939779 & 0.828525 & 5.013987 \\
937.516 & 8  & 8  & 3 & 3.915058 & 0.858394 & 5.048600 \\
935.937 & 8  & 8  & 4 & 3.878471 & 0.902459 & 5.099335 \\
933.826 & 8  & 8  & 5 & 3.828821 & 0.961885 & 5.167540 \\
931.122 & 8  & 8  & 6 & 3.763547 & 1.039360 & 5.256011 \\
927.742 & 8  & 8  & 7 & 3.679575 & 1.138211 & -        \\
940.836 & 9  & 9  & 1 & 4.594879 & 0.747287 & 5.512472 \\
940.228 & 9  & 9  & 2 & 4.581997 & 0.763583 & 5.531898 \\
939.198 & 9  & 9  & 3 & 4.559893 & 0.791531 & 5.565008 \\
937.699 & 9  & 9  & 4 & 4.527384 & 0.831969 & 5.612952 \\
935.707 & 9  & 9  & 5 & 4.482978 & 0.887018 & 5.677725 \\
933.157 & 9  & 9  & 6 & 4.424421 & 0.958828 & 5.761786 \\
929.970 & 9  & 9  & 7 & 4.348961 & 1.050521 & 5.868273 \\
926.046 & 9  & 9  & 8 & 4.252680 & 1.166164 & -        \\
942.420 & 10 & 10 & 1 & 5.231649 & 0.681503 & 6.023669 \\
941.851 & 10 & 10 & 2 & 5.220411 & 0.696434 & 6.042017 \\
940.866 & 10 & 10 & 3 & 5.201418 & 0.721263 & 6.072684 \\
939.479 & 10 & 10 & 4 & 5.172636 & 0.759001 & 6.118368 \\
937.612 & 10 & 10 & 5 & 5.133494 & 0.809481 & 6.179461 \\
935.221 & 10 & 10 & 6 & 5.081852 & 0.875368 & 6.258671 \\
932.235 & 10 & 10 & 7 & 5.015021 & 0.959566 & 6.359053 \\
928.558 & 10 & 10 & 8 & 4.929483 & 1.065868 & 6.484720 \\
924.070 & 10 & 10 & 9 & 4.820644 & 1.199163 & -        \\
\hline
\end{tabular*}
\end{table*}
\endgroup

\bibliography{references}